\documentclass[twocolumn,pra,twocolumn,superscriptaddress]{revtex4}
\usepackage{color}
\usepackage{amsmath}
\usepackage{amssymb}
\usepackage{graphicx}
\usepackage{tikz}

\usepackage{float}
\usepackage{bbm}

\usepackage{epsfig} 
\usepackage{verbatim}
\usepackage{array}
\usepackage{setspace}

\usepackage[unicode=true,
 bookmarks=true,bookmarksnumbered=false,bookmarksopen=false,
 breaklinks=false,pdfborder={0 0 1},backref=false,colorlinks=true]
 {hyperref}
\hypersetup{
 linkcolor=magenta, urlcolor=blue, citecolor=blue, pdfstartview={FitH}, hyperfootnotes=false, unicode=true}

\makeatletter
\@ifundefined{textcolor}{}
{%
 \definecolor{BLACK}{gray}{0}
 \definecolor{WHITE}{gray}{1}
 \definecolor{RED}{rgb}{1,0,0}
 \definecolor{GREEN}{rgb}{0,1,0}
 \definecolor{BLUE}{rgb}{0,0,1}
 \definecolor{CYAN}{cmyk}{1,0,0,0}
 \definecolor{MAGENTA}{cmyk}{0,1,0,0}
 \definecolor{YELLOW}{cmyk}{0,0,1,0}
}

\usepackage{amsfonts}\usepackage{tabularx}\usepackage{dcolumn}\usepackage{bm}\usepackage{graphicx}\usepackage{epstopdf}

\hypersetup{urlcolor=blue}

\makeatother
\newcolumntype{C}[1]{>{\centering\arraybackslash$}p{#1}<{$}}

\begin{document}

\title{Spin-qubit noise spectroscopy from randomized benchmarking by supervised learning}

\author{Chengxian Zhang}
\affiliation{Department of Physics, City University of Hong Kong, Tat Chee Avenue, Kowloon, Hong Kong SAR, China, and City University of Hong Kong Shenzhen Research Institute, Shenzhen, Guangdong 518057, China}
\affiliation{Guangdong Provincial Key Laboratory of Quantum Engineering and Quantum Materials, School of Physics and Telecommunication Engineering, South China Normal University, Guangzhou 510006, China}
\author{Xin Wang}
\email{x.wang@cityu.edu.hk}
\affiliation{Department of Physics, City University of Hong Kong, Tat Chee Avenue, Kowloon, Hong Kong SAR, China, and City University of Hong Kong Shenzhen Research Institute, Shenzhen, Guangdong 518057, China}
\date{\today}

\begin{abstract}
We demonstrate a method to obtain the spectra of $1/f$ noises in spin-qubit devices from randomized benchmarking, assisted by supervised learning. The noise exponent, which indicates the correlation within the noise, is determined by training a double-layer neural network with the ratio between the randomized benchmarking results of pulse sequences that correct noise and not. After the training is completed, the neural network is able to predict the exponent within an absolute error of about 0.05, comparable with existing methods. The noise amplitude is then evaluated by training another neural network with the decaying fidelity of randomized benchmarking results from the uncorrected sequences. The relative error for the prediction of the noise amplitude is as low as 5\% provided that the noise exponent is known. Our results suggest that the neural network is capable of predicting noise spectra from randomized benchmarking, which can be an alternative method to measure noise spectra in spin-qubit devices.
\end{abstract}

\maketitle

\section{Introduction}

Noises pose a major challenge to the physical realization of quantum computing. In quantum-dot spin qubits, decoherence is mainly caused by two sources of noises: nuclear noise and charge noise \cite{Kuhlmann.13}. The nuclear noise (or Overhauser noise) arises from, for example, the hyperfine coupling between the spin of the quantum-dot electron and the nuclear spins surrounding it \cite{Koppens.05, Reilly.08, Cywinski.09, Barnes.12,Chekhovich.13}. On the other hand, the charge noise is caused by unintentionally deposited impurities in the fabrication process of the samples, which an electron can hop on and off randomly \cite{Hu.06, Culcer.09, Nguyen.11}. To combat these noises, dynamically corrected gates 
(DCGs) \cite{Khodjasteh.10, Grace.12, Green.12, Wang.12, Kestner.13, Hickman.13, Seti.14} have been developed for various realizations of spin qubits, some of which have been experimentally demonstrated \cite{Rong.14}. Nevertheless, DCGs are mostly developed under the static noise approximation, i.e. the noises are assumed to be slowly varying as compared to the typical gate operation time. Therefore, the effectiveness of reducing noise for the DCGs depends critically on the correlation within the noise or, in the context of $1/f$ noise (with frequency spectrum $1/\omega^\alpha$) \cite{Castelano.16}, the ``exponent'' $\alpha$ \cite{Wang.14a, Yang.16}. When $\alpha$ is large, the noise is concentrated in low frequencies and DCGs are most effective. On the contrary, when $\alpha$ is small, the noise is white-like and DCGs would not offer improvements on the control quality. In a related context, it has been shown that the characteristics of noises play an important role in the optimal control theory, and robust control is typically limited to certain spectral regimes of the noises \cite{hocker.14}. These studies suggest that the characterization of noises and, in particular, the interplay between noises and gates is key to successful implementation of quantum computing \cite{Peddibhotla.13,Hansom.14} in quantum-dot spin systems.

Various methods have been used to measure the noise spectra using, for example, optical techniques \cite{Urbaszek.13, Glasenapp.14,Sinitsyn.16}, and dynamical-decoupling-based methods \cite{alvarez.11, Norris.16, Luke.17,Krzywda.18}. For semiconductor quantum dot systems, the exponent of the $1/f$ noise, $\alpha$, can be obtained through a scaling of the qubit's response to different orders of dynamical decoupling sequences \cite{Medford.12,Dial.13}. Specifically, the exponent for nuclear noise in a quantum-dot device has been estimated to be $2.5\sim2.6$ \cite{Rudner.11,Medford.12}, while that for charge noise is around $0.7\sim0.8$ \cite{Kuhlmann.13,Dial.13}. On the other hand, Randomized Benchmarking (RB) has been a tool ubiquitously used to understand the interplay between noises and gates, and in particular to evaluate the performance of the DCGs in both experimental and the theoretical studies \cite{Yuge.11,Watson.18,Kawakami.16}. In numerical simulations of RB, a sequence of gates, randomly drawn from the Clifford group, evolves under certain type of noises \cite{Magesan.12}. Exponential fitting of the result averaged over many runs gives the average error per gate. These studies have been useful to understand the effectiveness of DCGs undergoing $1/f$ noise with different exponents \cite{Zhang.17,Throckmorton.17}. For example, it has been found that DCGs would be effective in reducing noise in singlet-triplet qubits when the $1/f$ noise exponent is greater than 1 \cite{Yang.16}, while the threshold for exchange-only qubits is about 1.5 \cite{Zhang.16}. Nevertheless, the reverse problem, i.e.~obtaining the noise spectra from RB results has been difficult. 

The development of machine learning \cite{Mehta.18, Mohri.12} and especially supervised learning \cite{NielsenML} has provided a viable way to solve the problem. Machine learning is a set of techniques allowing data analysis or optimization in a way much more efficient than many enumerative methods previously known  \cite{Jordan.15,LeCun.15}. These techniques have been recently applied to various fields in physics with numerous success \cite{Biswas.13, Kalinin.15, Biamonte.16, Reddy.16, Schoenholz.16, WangLei.16,Nieuwenburg.17, Carleo.17,Bukov.18}.  In supervised  learning, a neural network is trained using  a large amount of data (including inputs and outputs). During the training process, the neurons adjust their parameters to match the input-output correspondence of the training set. Once the network is properly trained, the network is able to make predictions using inputs that are not in the training set. Using this feature, we have explored the possibility of constructing DCGs using supervised learning \cite{Yang.18}. It has been demonstrated that the trained neural network is capable of producing the composite pulse sequences that are as robust as the sequences known in the literature \cite{Wang.12,Kestner.13}. The trained network also enables interchanging the inputs and outputs, which allows us to estimate the noise spectra from RB results. In this paper, we demonstrate a method to measure the noise spectrum from RB employing supervised learning. The key to measure the noise spectrum is to determine the two parameters, namely, the noise exponent $\alpha$ and the noise amplitude $A$. We first train the neural network to predict the noise exponent $\alpha$ by feeding the ratios of the fidelities between the corrected and uncorrected sequences obtained from RB. This idea arises from the fact that the noise exponent $\alpha$ is key to the efficacy of DCGs \cite{Wang.14a, Yang.16, Zhang.16}. We found that a properly trained neural network is able to predict $\alpha$ within an error of about $\pm0.05$. On the other hand, there is a positive correlation between the noise amplitude $A$ and the decaying fidelity provided that the exponent $\alpha$ is fixed. Once $\alpha$ is known, the noise amplitude $A$ can be determined from how fast the fidelity decays for RB results of uncorrected sequences. We are able to predict $A$ with a relative error of $\sim5\%$.

This paper is organized as follows. In Sec.~\ref{sec:model}, we present the model and methods used in this work. Results are presented and explained in Sec.~\ref{sec:res}. We conclude in Sec.~\ref{sec:conclusion}.

\section{Model and Methods}
\label{sec:model}

\begin{figure}
	\includegraphics[width=0.8\columnwidth]{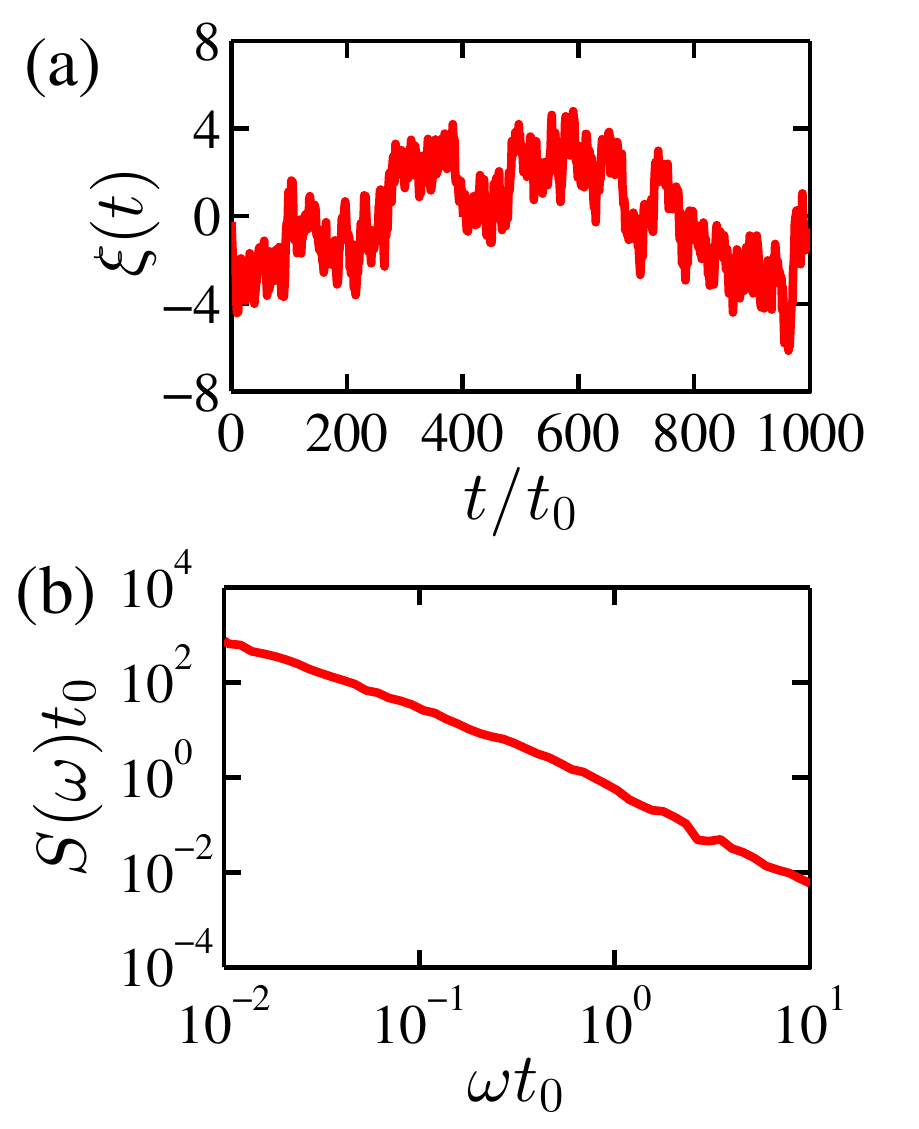}
	\caption{An example of the $1/f$ noise used in the simulation. (a) Noise as a function of time. (b) The power spectral density corresponding to the noise shown in (a). $S(\omega)=A/(\omega t_{0})^{1.5}$. The noise amplitude A has been scaled so that  $S(\omega=1/t_{0})\approx 1/t_{0}$, and $t_0$ is an arbitrary time unit.}
\label{fig:1}	
\end{figure}

Noises in semiconductor quantum-dot systems are commonly modeled by $1/f$ type \cite{Rudner.11,Medford.12,Kuhlmann.13,Dial.13,Yoneda.18,Kawakami.16}, which we consider in this work. Namely, their power spectral densities have the form
\begin{equation}
S(\omega) = A/(\omega t_0)^{\alpha},
\end{equation}
where $A$ denotes the noise amplitude, the exponent $\alpha$ indicates the correlation within the noise, and $t_0$ is an arbitrary time unit. For small  $\alpha$, the noise is close to white noise;  when $\alpha$ is large, the noise is concentrated in low frequencies. Therefore, the key of noise spectroscopy is the determination of the two parameters, $A$ and $\alpha$. Traditionally, one generates $1/f$ noises by summing random telegraph signals \cite{Wang.14a, Castelano.16}. Nevertheless, we have found that this method can only reliably generate noises with exponent $1/2\lesssim\alpha\lesssim3/2$. In this work, we use an alternative method based on the inverse Fourier transform of noises in the frequency domain, capable to extend the range of the noise exponent to $0\leq\alpha\leq3$ \cite{Yang.16}. Fig.~\ref{fig:1} shows an example of $1/f$ noise with $\alpha=1.5$. Note that the frequency has a dimension of the reciprocal of time (equivalent to energy if $\hbar=1$), therefore it is multiplied by an arbitrary time unit $t_0$ to make it dimensionless. Both the power spectral density $S(\omega)$ and the noise amplitude have a dimension of energy (the reciprocal of time if $\hbar=1$), therefore in later discussions they are also sometimes multiplied by $t_0$ when dimensionless quantities are desired. Fig.~\ref{fig:1}(a) depicts the noise as a function of time, $\xi(t)$. Fig.~\ref{fig:1}(b) shows the corresponding dimensionless power spectral density, $S(\omega)t_0$, which is close to a straight line under the log scale.
	
\begin{figure}
	\includegraphics[width=1\columnwidth]{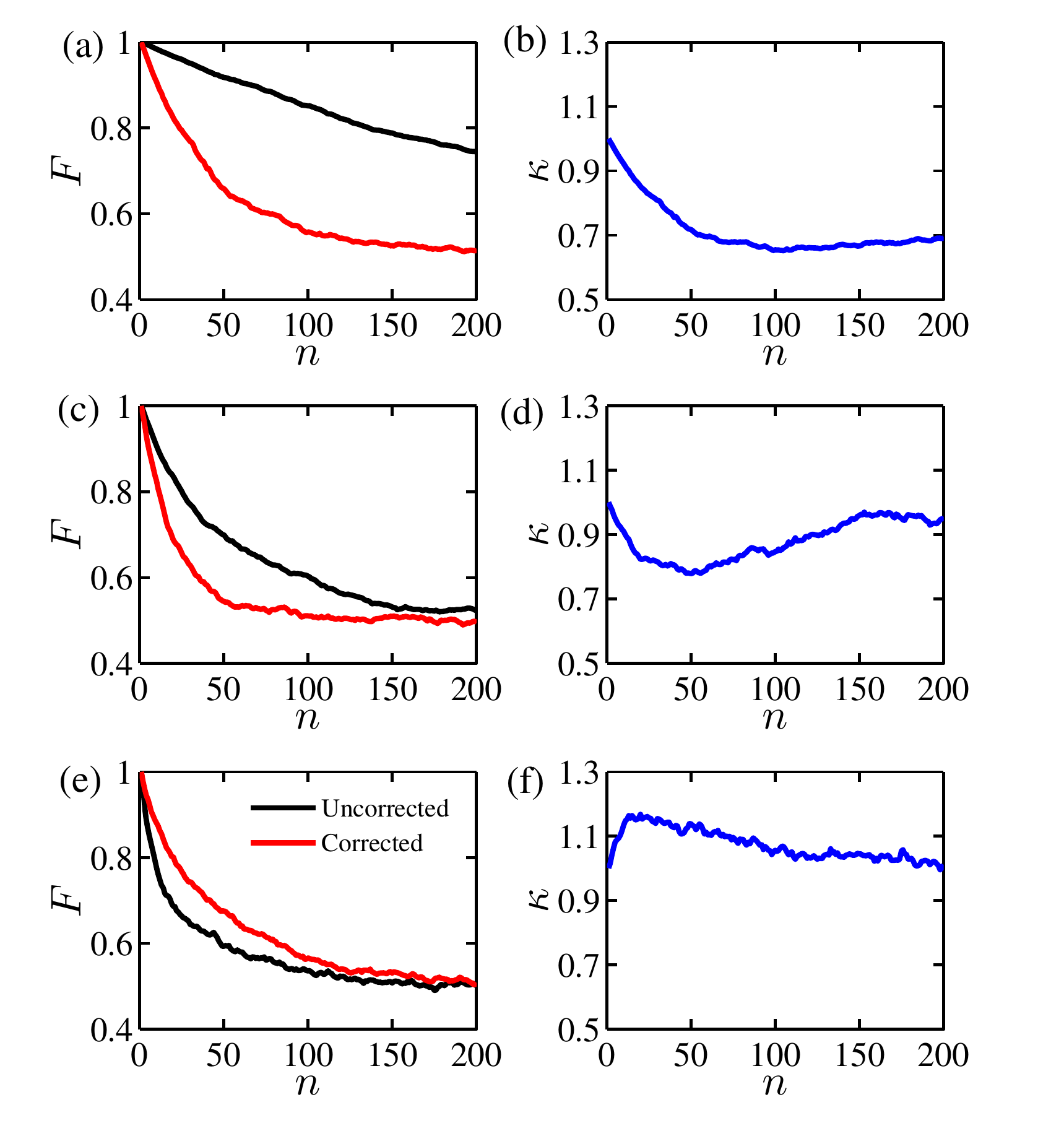}
	\caption{Randomized benchmarking results of single-qubit Clifford gates subject to $1/f$ noise with different  exponents. (a), (c), (e): The fidelities for uncorrected and corrected gates are shown as black and red/gray lines respectively. (b), (d), (f): The ratios between the corrected and uncorrected sequences corresponds to the results shown on the left. The noise exponents are: (a), (b) $\alpha=0$; (c), (d) $\alpha=1$; (e), (f) $\alpha=1.5$. The noise amplitude is fixed by $At_{0}=10^{-3}$. }
	\label{fig:2}
\end{figure}

A workhorse to understand the performance of quantum gates subject to time-dependent noises is RB which can be simulated on a computer. In simulated (single-qubit) RB,  sequences of quantum gates randomly drawn from the 24 single-qubit Clifford gates are executed in presence of noises. Under the assumption that the gate errors are uncorrelated, the fidelity $F$ decays in an exponential form
\begin{equation}
F=\frac{1}{2}\left(1+e^{-\gamma n}\right),
\end{equation}
where $n$ is the number of gates and $\gamma$ is a parameter closely related to the average error per gate. Essentially, RB takes noises and gates as inputs and the decaying fidelities as outputs. The left column of Fig.~\ref{fig:2} [panels (a), (c) and (e)] shows typical results of RB using gates for a singlet-triplet qubit as explained in \cite{Wang.14a}. The black lines are for ``uncorrected'' quantum gates, i.e. gates that are not immune to noises. The red/gray lines are for ``corrected'' gates, which refer to the {\sc supcode} sequences \cite{Wang.12,Kestner.13} that are resilient to both nuclear and charge noises. For simplicity, in this work we consider the case that the nuclear and charge noises have the same spectrum, and we shall generically refer to them as ``noises''. 

For noises with a given spectra and a set of well-defined gates, it is straightforward to run RB simulations and obtain the average error per gate by exponential fitting. Nevertheless, the inverse problem is complicated for two reasons. Firstly, RB is a statistic process during which certain information is lost or suppressed, so it would be hard to recover the noise spectra from just the averaged fidelity. Secondly, the noise spectrum contains two key parameters, the noise amplitude $A$ and exponent $\alpha$, both of which contribute to gate errors at the same time and are hard to distinguish. In this work, we shall use the neural network and supervised learning, along with our existing understanding of the interplay between quantum gates and noises, to overcome these difficulties. We shall show that upon judicious training of two neural networks, we will be able to determine the noise amplitude $A$ to a precision of about 5\% (relative error) and the exponent $\alpha$ to about $\pm0.05$ (absolute error) from given inputs of RB data. 

Supervised learning \cite{Mohri.12,NielsenML} is a branch of machine learning which uses a large amount of data to train a neural network. During the training process, the network ``learns'' from a set of data (``training set'') with known inputs and outputs. Essentially, the parameters governing the network (such as weights and biases) are optimized iteratively such that the outputs of network fit better and better to the known output from the training set. Once the network is properly trained (i.e. the outputs from the network fit known ones to the desired precision),  it becomes capable to predict unknown outputs from given inputs. A detailed description of supervised learning and its application in generating composite pulse sequences to control spin qubits can be found in \cite{Yang.18}.

Supervised learning provides a viable method to interchange the inputs and outputs of the RB process. Namely, one may train the neural network with the averaged fidelities (outputs from RB) as inputs and the parameters of noise spectra (inputs to RB) as outputs, and the network shall be able to predict the noise spectrum from a known RB result.  Nevertheless, since a noise spectrum contain two parameters, the amplitude $A$ and exponent $\alpha$, and they both contribute to the RB results, we must find a way to distinguish the two. On one hand, there is a positive correlation between the noise amplitude $A$ and the decaying fidelity provided that the exponent $\alpha$ is fixed. A large $A$ implies higher noise level, which leads to faster decay of the gate fidelity of the RB results. On the other hand, the noise exponent $\alpha$ is deeply involved in the efficacy of the noise-correcting composite pulse sequences. Since these sequences are developed with the static noise approximation, they should work  for $1/f^\alpha$ noises with larger $\alpha$ but not otherwise. This observation has been confirmed in our previous works on {\sc supcode} \cite{Wang.14a,Wang.14b,Yang.16}. For small $\alpha$ ($\alpha \lesssim\alpha_c$ where $\alpha_c$ is a critical value at which non-correcting sequences and DCGs perform comparably under noise \cite{Yang.16}), the noise is white-like, the corrected sequences perform worse than the uncorrected ones. For large $\alpha$ ($\alpha \gtrsim \alpha_c$), the noise concentrates more at low frequencies and the corrected sequences outperform the uncorrected ones. For {\sc supcode} and the singlet-triplet qubit, the critical value $\alpha_c$ is found to be around 1 \cite{Yang.16}. This observation can be quantified by the ratio between the fidelity of the two set of states, defined by the fidelity of the corrected gate sequence divided by that of the uncorrected one, and we denote it by $\kappa$. The right column of Fig.~\ref{fig:2} shows $\kappa$ corresponding to its counterparts in the left column. For Fig.~\ref{fig:2}(b) ($\alpha=0$), $\kappa$ is smaller than 1 and saturate to a value around 0.7. For Fig.~\ref{fig:2}(d) ($\alpha=1$), $\kappa$ fluctuates around 1. For Fig.~\ref{fig:2}(f) ($\alpha=1.5$), $\kappa$ remains greater than 1. Because the value of $\kappa$ is related to $\alpha$, one may use $\kappa$ as inputs to the neural network and $\alpha$ as the output, thereby determining $\alpha$.

Summarizing the understandings above, we use a two-step strategy to obtain the noise spectra. We first determine $\alpha$ by training a neural network with the ratio between the RB results of the corrected and uncorrected sequences as input, and $\alpha$ as the output. Once $\alpha$ is determined, we train another neural network with the RB results from either the corrected or uncorrected sequences (in this work the uncorrected ones are used) as inputs and $A$ as the output, with which $A$ is determined.

\section{Results}
\label{sec:res}

\begin{figure}
	\includegraphics[width=1\columnwidth]{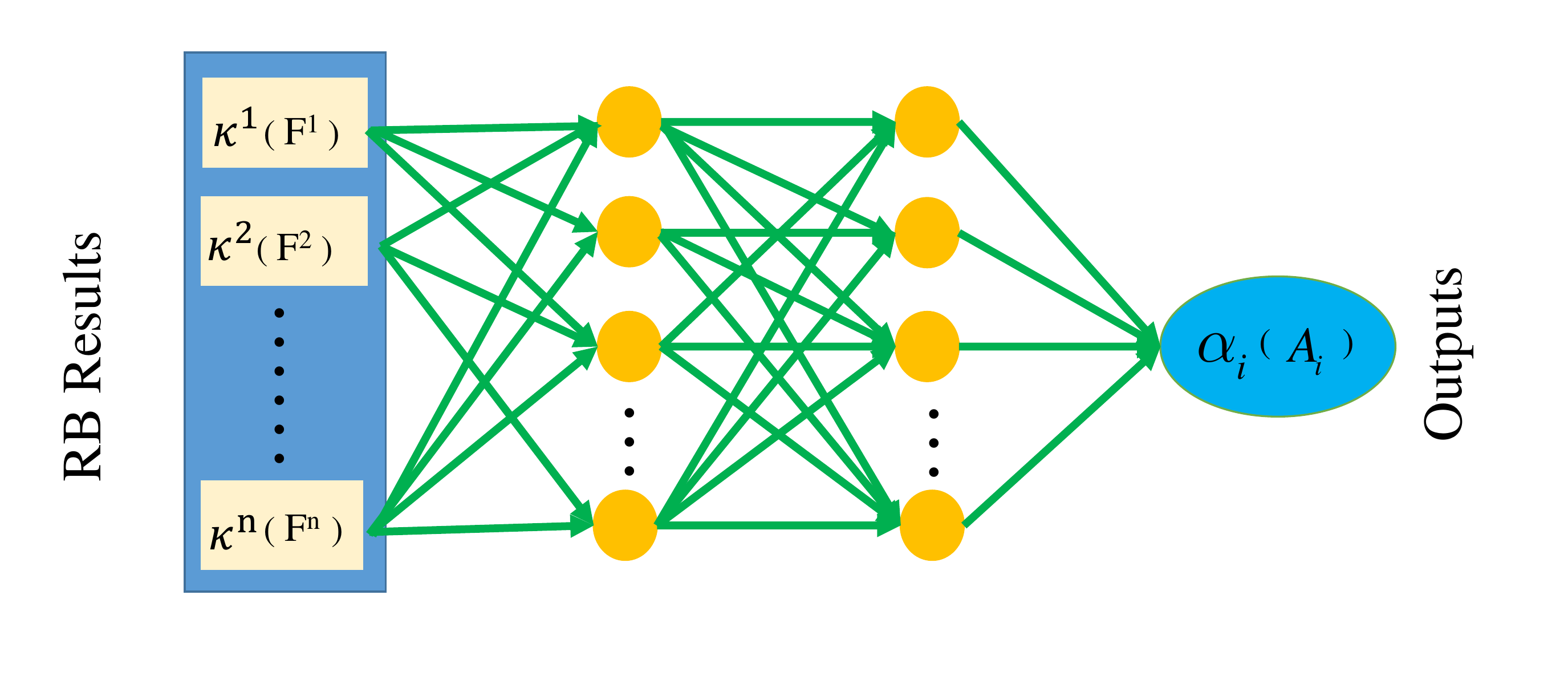}
	\caption{Schematics of the neural network used in this work. The inputs are given by RB results: $\kappa$ (corresponding to output of $\alpha$) and $F$ (corresponding to $A$). $\kappa$ and $F$ are discretized into  $\kappa^i$ and $F^i$ where $i=1,\ldots, N$ and $N=200$. The neural network contains two hidden layers with $N_{n}$ neurons each.}
	\label{fig:3}	
\end{figure} 

The neural network used in our work has two hidden layers with $N_{n}$ neurons each. Fig.~\ref{fig:3} schematically shows the structure. For the convenience of later discussions, the default parameters of the neural network are provided in Table~\ref{tablenet}.  One may also refer to Appendix \ref{appx:suplearning} for details on the procedure of supervised learning.

In determining $\alpha$, the inputs are RB ratios $\kappa$ between the corrected and uncorrected sequences as functions of the number of gates, which are read to the network as vectors $\kappa^i$ $(i=1,2,\ldots,N)$ where $N$ is the number of gates in a sequence used in the simulation. Similarly, in determining $A$ the inputs are RB results $F^i$ of the uncorrected sequences assuming that $\alpha$ is already known. In order to ensure convergence, each data point in the training set is an average of 200 RB simulation results with different gate sequences undergoing different noise realizations for a given noise spectra.  For determining $\alpha$, the training data points are obtained for different noise spectra with amplitude $\tilde{A}$ and $\tilde{\alpha}$ as follows: 50 points of different $\tilde{\alpha}$ are chosen uniformly on $\tilde{\alpha}\in (0,3)$, and 25 points of different $\tilde{A}$ are chosen such that $\log(\tilde{A}t_0)$ distribute uniformly on $\log(\tilde{A}t_0)\in ({-7},{-4})$. Here we note that the value of dimensionless noise amplitude, $At_0$ depends on the choice of $t_0$. In \cite{Medford.12}, the measured strength of the nuclear noise at $\alpha\approx2$ can be estimated to be $At_0\approx10^{-6}$ using $t_0=10$ ns.  With the same $t_0$,  we convert the charge noise data from \cite{Kawakami.16} to $At_0\approx10^{-8}$ at $\alpha\approx2$ \cite{Yang.16, Zhang.16}. In this work, we use $At_0=10^{-8}$ as the lower bound of the noise strength and $10^{-2}$ as the upper bound in determining the noise amplitude, $A$. However, for the training discussed here, we involve two variables $\alpha$ and $A$ with the main aim determining $\alpha$.  We therefore have chosen to reduce the range of $A$ considered to $10^{-7}\sim10^{-4}$ in order to keep the size of the training set tractable. Our method can certainly be straightforwardly extended to cover a wider range of $A$ with more data in the training set.

For each specific $\tilde{A}$ and $\tilde{\alpha}$, 8 data points (each of which is an average of 200 RB simulations and is obtained with independent RB runs) are generated. Therefore the training data set has $50\times25\times8=10000$ entries in total.  These data are used to train a neural network with training parameters including learning rate $\eta$, bin size $b$, and number of epochs $N_{e}$.  (One may refer to \cite{Yang.18} for a detailed explanation of the parameters) The default values of these parameters are provided in Table~\ref{tablenet} but when we use different values they are going to be specified.

For determining $A$, $\alpha$ is already known (and we have taken $\alpha=1.5$ as a representative point). The training data points are therefore obtained for noise with different amplitude $\tilde{A}$, chosen such that there are 400 points of $\log(\tilde{A}t_0)$ distributing uniformly on $\log(\tilde{A}t_0)\in ({-8},{-2})$. For each $\tilde{A}$, we generate 25 data points (each of which is an average of 200 RB simulations and is obtained with independent RB runs). The training data set for determining $A$ therefore has $400\times25=10000$ entries in total. These data are used to train a separate neural network in order to determine $A$, but the structure of the network is similar to that used to determine $\alpha$.

\begin{table}[]
  \begin{center}
    \caption{Default parameters of the neural network.}
      \label{tablenet}
    \begin{tabular}{p{6cm}rp{5cm}}
    \hline\hline
      Number of layers & 2\\
      Number of neurons in each layer $N_n$ & 50\\
      Size of the training data set $N_{\rm tr}$ & 5000\\
      Size of a data bin $b$ & 10\\
      Number of training epochs $N_e$ & 1000 \\
      Activation function $f(z)$ & $1/(1+e^{-z})$\\
      Learning rate $\eta$ & 0.005\\
      \hline\hline
    \end{tabular}
  \end{center}
\end{table}

\begin{figure}
	\includegraphics[width=1.1\columnwidth]{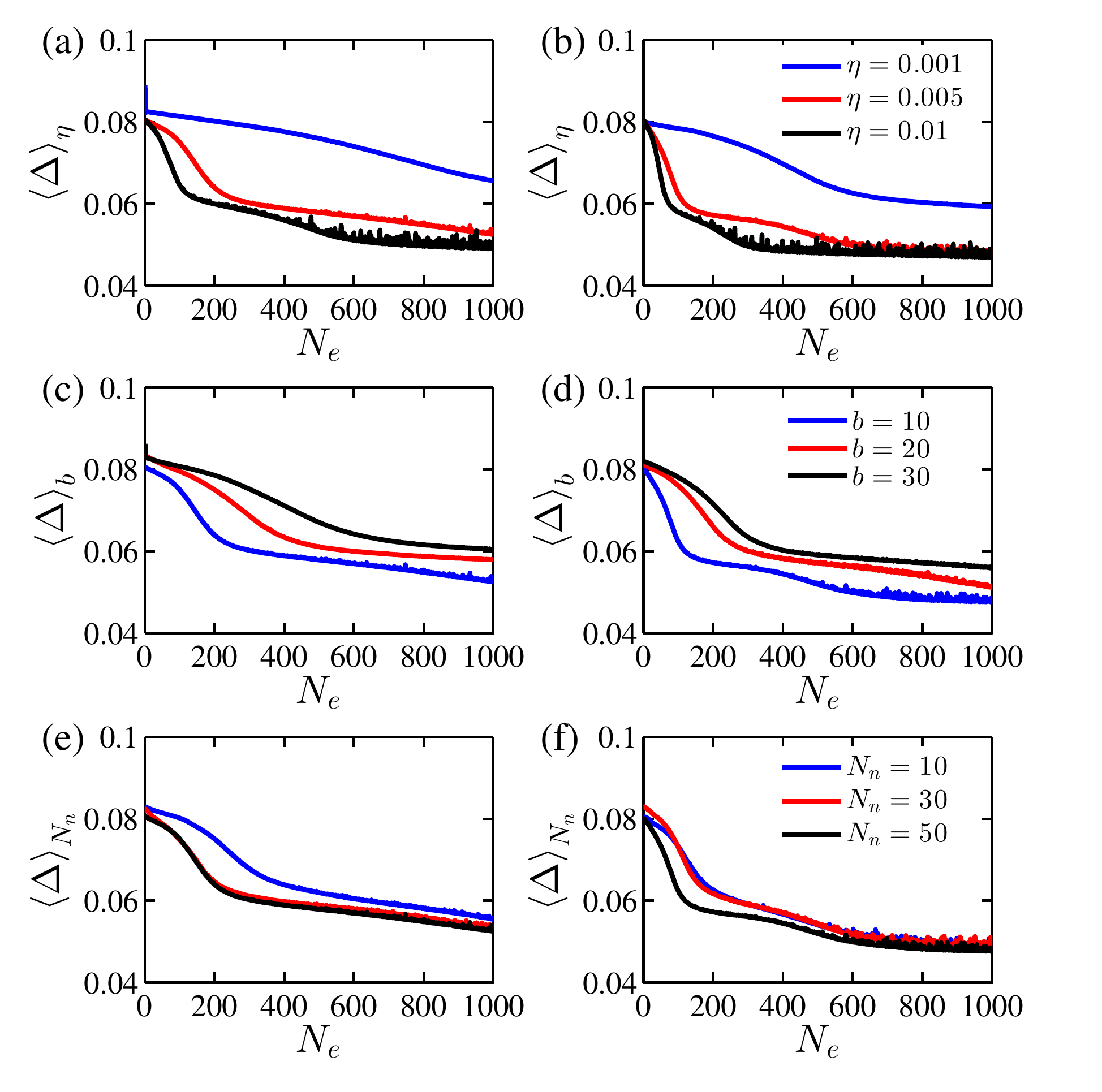}
	\caption{ The averaged error $\Delta$ for the noise exponent $\alpha$ as functions of the number of epochs. The left column [panels (a), (c) and (e)] corresponds to a training set with 5000 entries, while the right column [panels (b), (d) and (f)] corresponds to a training set with 10000 entries. For panels (a) and (b), the blue (upper) line, red (middle) line and black (bottom) line show results with learning rates $\eta=0.001$, 0.005 and 0.01 respectively, while other parameters are fixed at $b=10$ and $N_n=50$. For panels (c) and (d), the blue (bottom) line, red (middle) line and black (upper) line show results with bin size $b=10$, 20 and 30 respectively, while other parameters are fixed at $\eta=0.005$ and $N_{n}=50$. Lines in panels (e) and (f) show results with $N_{n}=10$, 30 and 50 respectively while other parameters are fixed at $\eta=0.005$ and $b=10$.}
    \label{fig:4}
\end{figure} 

In order to quantify the accuracy of $\alpha$ predicted by the neural network, we define the average error $\Delta=\frac{1}{n}\sum_{i=1}^{n}\left| \alpha-\alpha_{i} \right|$ where $\alpha$ is the known noise exponent and $\alpha_{i}$ the corresponding predicted one, and $n$ is the number of data points available for testing. In Fig.~\ref{fig:4}, we plot $\Delta$ as functions of the number of epochs for different training parameters, including the learning rate $\eta$, bin size $b$ and number of neurons in each layer $N_{n}$. The left column [panels (a), (c) and (e)] shows results with the size of training set $N_{\rm tr}=5000$, while the right column [panels (b), (d) and (f)] shows results for $N_{\rm tr}=10000$. It is clear that in all cases, results with a larger training set are better than ones with a smaller training set, as the errors of predicted $\alpha$ converge to a lower value in the right column as compared to the left one. 

Fig.~\ref{fig:4}(a) and (b) show results with different training rates, $\eta$. For small $\eta$, the error goes down with more epochs smoothly but slowly, while for larger $\eta$, the error goes down more steeply but is unstable, i.e. more oscillations can be seen. For $\eta=0.001$, the average error reduce to 0.06 for the case with $N_{\rm tr}=10000$, significantly larger than the other cases. Since the average errors for both $\eta=0.005$ and $\eta=0.01$   converge to 0.05 but that for $\eta=0.01$ is more unstable, we consider $\eta=0.005$ an appropriate value for the learning rate. Fig.~\ref{fig:4}(c) and (d) show results with different bin sizes. When the bin size is large (e.g, $b\ge20$),  the error goes down slower than the case with a smaller bin size. However, when the training data is noisy, a bin size which is too small could potentially lead to overfitting on the noisy details of the data. Since the results for $b=10$ do not show signs of overfitting, we consider $b=10$ appropriate for this training. Fig.~\ref{fig:4}(e) and (f) show results with different number of neurons in each hidden layer.  While increasing the number of neurons may improve the ability of the network to fit the training data, we see that for the case with $N_{\rm tr}=10000$, the difference is not large. For the case with $N_{\rm tr}=5000$, result for $N_n=10$ is notably worse than the other two cases, while the result for $N_n=50$ is the best with the lowest error. We therefore take $N_n=50$ as the default parameter of the neural network. Overall we have shown that with proper training, the neural network can predict $\alpha$ with an accuracy of about 0.05 (absolute error), which is either on par with or exceed existing methods based on dynamical decoupling \cite{Medford.12,Dial.13}.

\begin{figure}
	\includegraphics[width=1\columnwidth]{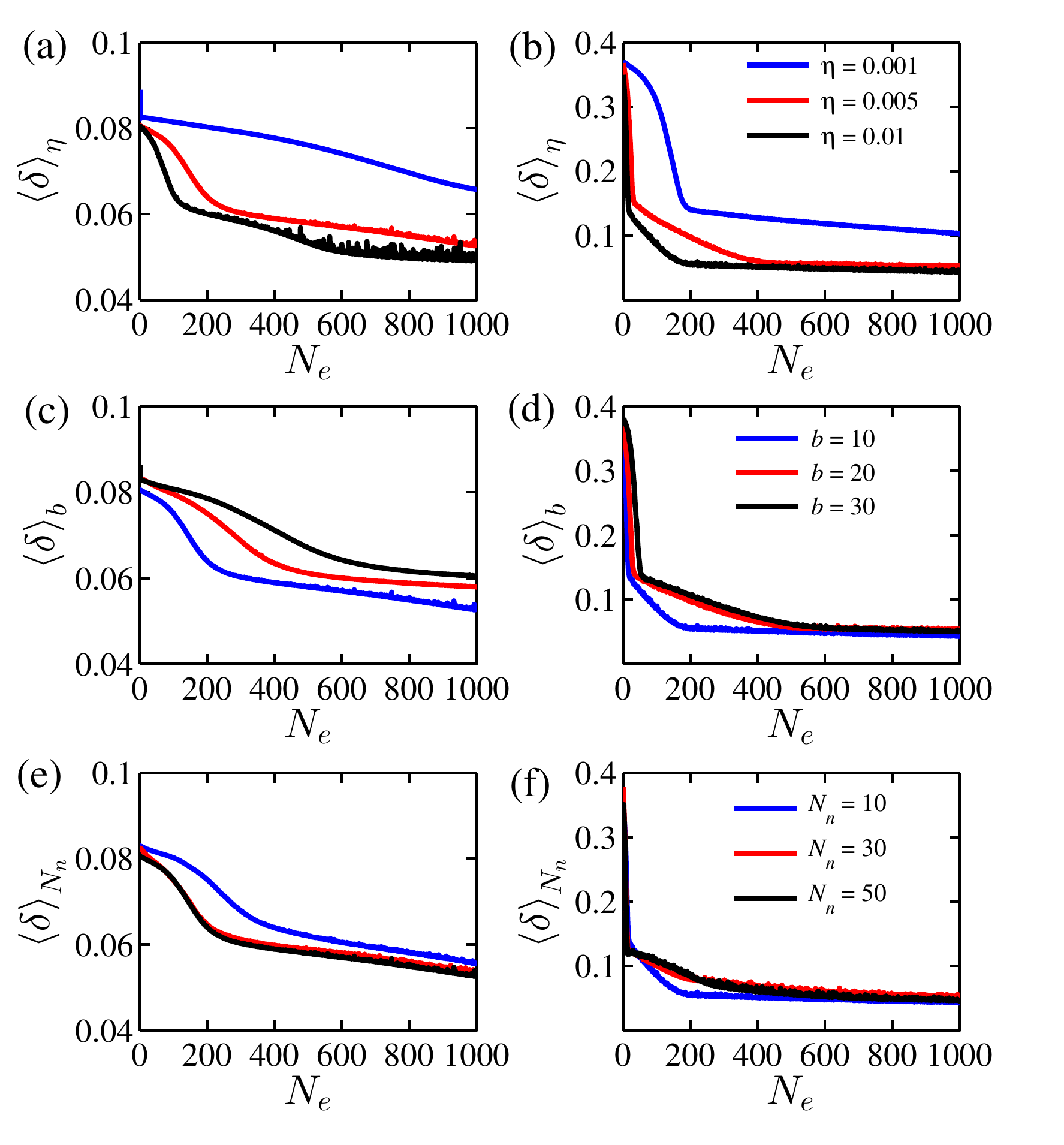}
	\caption{The (dimensionless) averaged relative error $\delta$ for the noise amplitude $At_{0}$ as functions of the number of epochs. The left column [panels (a), (c) and (e)] corresponds to a training set with 5000 entries, while the right column [panels (b), (d) and (f)] corresponds to a training set with 10000 entries. For panels (a) and (b), the blue (upper) line, red (middle) line and black (bottom) line show results with learning rates $\eta=0.001$, 0.005 and 0.01 respectively.
For panels (c) and (d), the blue (bottom) line, red (middle) line and black (upper) line show results with bin size $b=10$, 20 and 30 respectively. Lines in panels (e) and (f) show results with $N_{n}=10$, 30 and 50. Except where explicitly noted, the parameters are set at their defult values as in Table~\ref{tablenet}.}
	\label{fig:5}
\end{figure} 

Next we proceed to show the results of predicting the noise amplitude $A$ in Fig.~\ref{fig:5}. Since the value of $A$ may span several orders of magnitude, and $A$ is always nonzero, we consider the averaged \emph{relative} error of $A$, defined as $\delta=\frac{1}{n}\sum_{i=1}^{n}\left| \frac{A-A_{i}}{A} \right|$ where $A$ is the known noise amplitude, and $A_{i}$ the predicted one. Similar to the previous case of determining $\alpha$, we prepare two training data sets, one with $N_{\rm tr}=5000$ (shown as the left column of Fig.~\ref{fig:5}) and one with $N_{\rm tr}=10000$ (right column of Fig.~\ref{fig:5}). Qualitatively, the results are similar to Fig.~\ref{fig:4}. The saturated error is smaller for the network trained by a larger training set. Fig.~\ref{fig:5}(a) and (b) show results with different learning rates $\eta$. Again, $\eta=0.001$ is too small and the error is not reduced efficiently. For $N_{\rm tr}=10000$ results [Fig.~\ref{fig:5}(b)], both lines for $\eta=0.005$ and $\eta=0.01$ converge to the relative error level around 5\%. Fig.~\ref{fig:5}(c) and (d) show results with different bin sizes $b$. While different bin sizes make considerable differences in Fig.~\ref{fig:5}(c) ($N_{\rm tr}=5000$), they all converge to the relative error about 5\% in Fig.~\ref{fig:5}(d) ($N_{\rm tr}=10000$). The behavior is similar in Fig.~\ref{fig:5}(e) and (f) where results for different number of neurons in each hidden layer are displayed. Again, when the training set has $N_{\rm tr}=10000$ data points, the relative error for all cases converge to around 5\%. Overall, we have shown that a properly trained neural network can predict the noise amplitude of a device to an accuracy of about 5\%, provided that the noise exponent $\alpha$ is known.

\section{Conclusions}
\label{sec:conclusion}

To conclude, we have shown that judiciously trained neural networks can be used to measure the spectrum of $1/f$ noise. We determine the noise exponent $\alpha$ and the amplitude $A$ through a two-step process. Firstly, we perform RB for DCGs and gates not immune to noise. The ratio between the RB fidelities of the two sets of gates are fed to a double-layer neural network as inputs and the known noise exponents as the outputs for training. We found that after the neural network is properly trained, it can predict the noise exponent with an absolute error about $\pm0.05$. Then we perform RB with only the uncorrected pulse sequences and use their decaying fidelities as inputs to another neural network, while the noise amplitudes are outputs. We then show that provided the noise exponent is known, the neural network can predict the noise amplitude with a relative error of $\sim5\%$. 
Overall, we have shown that supervised learning in combination with RB provides an alternative method to measure the noise spectrum in quantum-dot devices. Lastly, we note that the $1/f$ noise we considered in our study is a rather simplified case. Evaluating noises with more complicated spectra is in principle possible with RB and machine learning, but would require more complex neural networks as well as more extensive training processes.

This work is supported by the 
Research Grants Council of the Hong Kong Special Administrative Region, China (No.~CityU 11303617, CityU 11304018), the National Natural Science Foundation of China (No.~11874312, 11604277), and the Guangdong Innovative and Entrepreneurial Research Team Program (No.~2016ZT06D348).

\appendix
\renewcommand{\theequation}{A-\arabic{equation}}
\setcounter{equation}{0}

\section{Brief description of supervised learning}\label{appx:suplearning}

\begin{figure}
	\includegraphics[width=1\columnwidth]{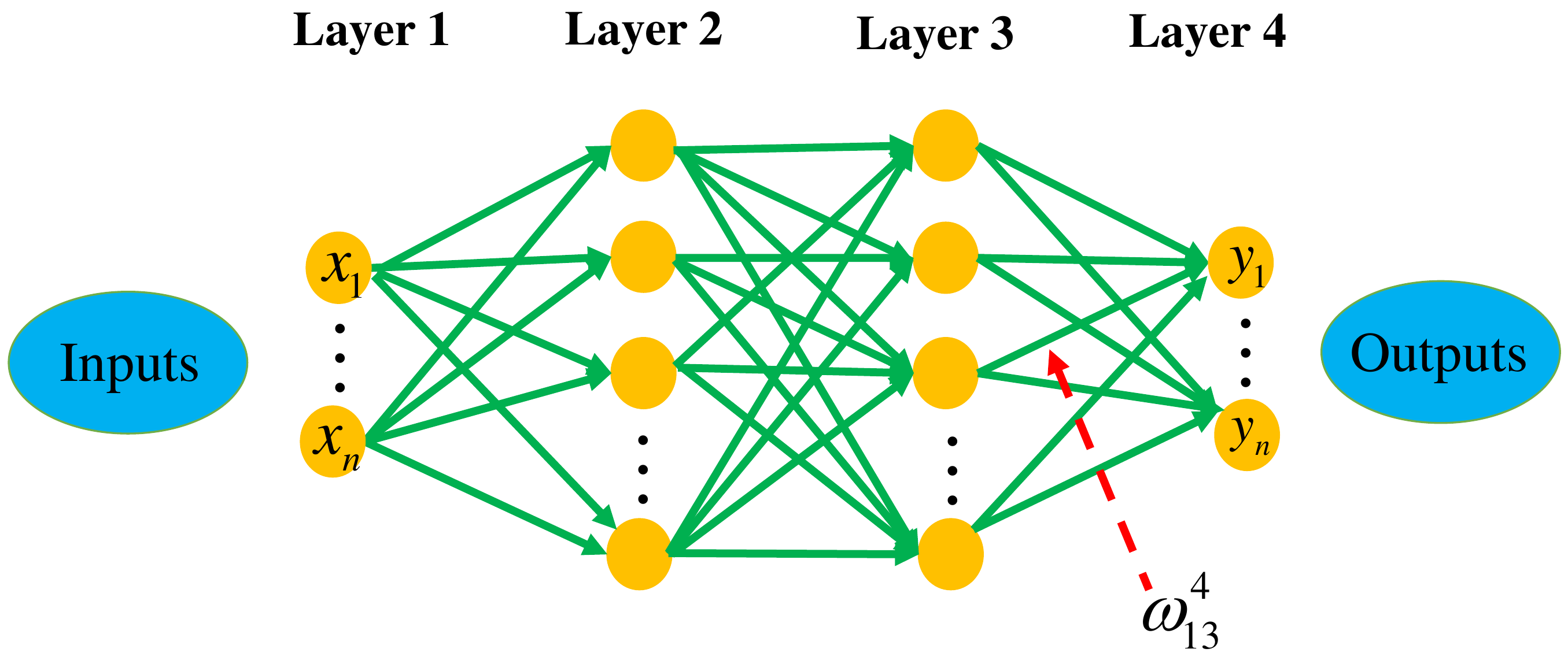}
	\caption{Schematic of a neural network used to describe the back-propagation algorithm. $\omega_{jk}^{l}$ denotes the weight of $j^{\rm th}$ neuron in the $l^{\rm th}$ layer to the $k^{\rm th}$ neuron in the $(l-1)^{\rm th}$ layer.}
	\label{fig:appen_network}
\end{figure}

In this section, we give a brief description of supervised learning as well as the  back-propagation algorithm used to update the neural network in the training process. This section largely follows \cite{NielsenML, Yang.18}.

An example of the neural network is shown in Fig.~\ref{fig:appen_network}. The first layer is for the inputs, while the last layer is for outputs. Each neuron has a bias and a set of weights: $\omega_{jk}^{l}$ (weight) connects the $j^{\rm th}$ neuron in the $l^{\rm th}$ layer to the $k^{\rm th}$ neuron in the $(l-1)^{\rm th}$ layer, while $d_{j}^{l}$ denotes the bias for the $j^{\rm th}$ neuron in $l^{\rm th}$ layer. The information then propagates through the network via a ``weighted input'' of the $j^{\rm th}$ neuron in the $l^{\rm th}$ layer from the $(l-1)^{\rm th}$ layer
\begin{equation}
z_{j}^{l}=\sum_{k}\omega_{jk}^{l}a_{k}^{l-1}+d_{j}^{l}.\label{lthz}
\end{equation}
Note that the summation over $k$ covers all neurons in the $(l-1)^{\rm th}$ layer. The output of the $j^{\rm th}$ neuron in the $l^{\rm th}$ layer is given via an activation function $f(z)$ as
\begin{equation}
a_{j}^{l}=f(z_{j}^{l}).\label{ltha} 
\end{equation}
In this work we take $f(z)=1/(1+e^{-z})$, but other forms are certainly allowed.

Consider a network with a total of $L$ layers which has not yet been well trained. The outputs from the network, $a_j^L$ (where $L$ denotes the last layer---the output layer), are different from the desired ones, which we denote by $y_j$. In this case, we evaluate the difference between them, based on which we modify the weights and bias of the network so that it will fit better to the data in the next run. This is essentially the training process. The difference between the actual prediction from the network and the desired outputs is evaluated using a cost function
\begin{equation}
C_{m}=\frac{1}{2}\sum_{j}(y_{j}-a_{j}^{L}),\label{costfunc}
\end{equation}
where $m$ denotes a specific training example. Averaging over the entire training set containing $N_{\rm tr}$ data, the cost function is
\begin{equation}
C=\frac{1}{N_{\rm tr}}\sum_{m}C_{m}.
\label{sumcostfunc}
\end{equation}

We then update the weights and bias of the network based on the back-propagation algorithm. The back-propagation algorithm is essentially a type of gradient descent method aiming at finding out the partial derivatives $\partial C/\partial \omega_{jk}^{l}$ and $\partial C /\partial d_{j}^l$ at layer $l$ such that the weight and bias can be modified in the $(l-1)^{\rm th}$ layer in order to better fit the training data. To compute the derivatives, we start with a quantity that denotes the error in the $j^{\rm th}$ neuron of the $l^{\rm th}$ layer, $\delta_{j}^{l}$, as
\begin{equation}
\delta_{j}^{l}=\frac{\partial C}{\partial z_{j}^{l}}.\label{deltaj}
\end{equation}
It can also be shown that $\partial C / \partial \omega_{jk}^{l}=a_{k}^{l-1}\delta_{j}^{l}$. Therefore the remaining task amounts to calculating $\delta_{j}^l$. 

We start from the output layer. From Eq.~\eqref{deltaj}, we have
\begin{equation}
\delta_{j}^{L}= \frac{\partial C}{\partial z_{j}^{L}}= \frac{\partial C}{\partial a_{j}^{L}} \frac{\partial a_{j}^{L}}{\partial z_{j}^{L}}=\frac{\partial C}{\partial a_{j}^{L}} f'(z_{j}^{L}).\label{deltajL}
\end{equation}
We also have
\begin{equation}
\delta_{j}^{l}= \frac{\partial C}{\partial z_{j}^{l}}=\sum_{k} \frac{\partial C}{\partial z_{k}^{l+1}} \frac{\partial z_{k}^{l+1}}{\partial z_{j}^{l}}=\sum_{k}\delta_{k}^{l+1}\frac{\partial z_{k}^{l+1}}{\partial z_{j}^{l}}.\label{deltajL2}
\end{equation}
On the other hand, it is straightforward to obtain
\begin{equation}
\frac{\partial z_{k}^{l+1}}{\partial z_{j}^{l}}=\omega_{kj}^{l+1}f'(z_{j}^{l}).\label{partialzk}
\end{equation}
Therefore, Eq.~\eqref{deltajL2} can be rewritten as 
\begin{equation}
\delta_{j}^{l}=f'(z_{j}^{l})\sum_{k}\omega_{kj}^{l+1}\delta_{k}^{l+1}.\label{deltajl2}
\end{equation}

Eq.~\eqref{deltajl2} is therefore a recursion relation that relates the error of the $(l+1)^{\rm th}$ layer and that of the $l^{\rm th}$ layer. These error values are then used to update the weights and bias of the network:
\begin{equation}
\omega_{jk}^{l}\rightarrow \omega_{jk}^{l}-\eta\partial C/\partial\omega_{jk}^{l},\label{modifiedw}
\end{equation}
\begin{equation}
d_{j}^l\rightarrow d_{j}^l-\eta\partial C/\partial d_{j}^l.\label{modifiedb}
\end{equation}
so that the cost function is reduced. Here, $\eta$ is the learning rate which determines the extent to which the update is done. One has to choose an $\eta$ appropriate to the individual situation so that the training process can efficiently converge.

The training process is summarized as follows:
\begin{enumerate}
\item For each given training example (including inputs $x_j$ and outputs $y_j$), set the activation for the first layer as $a_j^{1}=x_j$.
\item Feed forward according to Eq.~\eqref{ltha} so as to get $a_j^{l}$ for each hidden layer and, eventually, $a_j^{L}$ of the output layer.
\item Calculate the errors $\delta_{j}^{L}$ of the output layer with respect to the known outputs $y_j$ according to Eq.~\eqref{deltajL}.
\item Back-propogate the errors according to the recursion formula in Eq.~\eqref{deltajl2}.
\item Calculate the the partial derivatives $\partial C/\partial \omega_{jk}^{l}$ and $\partial C/\partial d_{j}^l$.
\item Set the proper learning rate $\eta$ to renew the weight and the bias throughout the network.
\item Repeat the entire procedure with other training data until the cost function is below certain threshold, or after a preset number of iterations.
\end{enumerate}

In practice, one sometimes re-bin the training set into batches with size $b$ (i.e., each bin contains an average of $b$ training data), in order to minimize the random error existing in the data. A training that finishes up using all data is called one epoch. After one epoch, one re-shuffles the training set and re-bin the data obtaining $N_{\rm tr}/b$ bins different than the previous epoch, and trains the network again. One typically needs at least hundreds of epochs in the entire training process to ensure convergence, and in our case we run each training up to 1000 epochs. The number of epochs has been labeled as $N_e$ in the paper.

\section{Results from neural networks with more layers}\label{4_hidden_layer}

\begin{figure}
	\includegraphics[width=0.85\columnwidth]{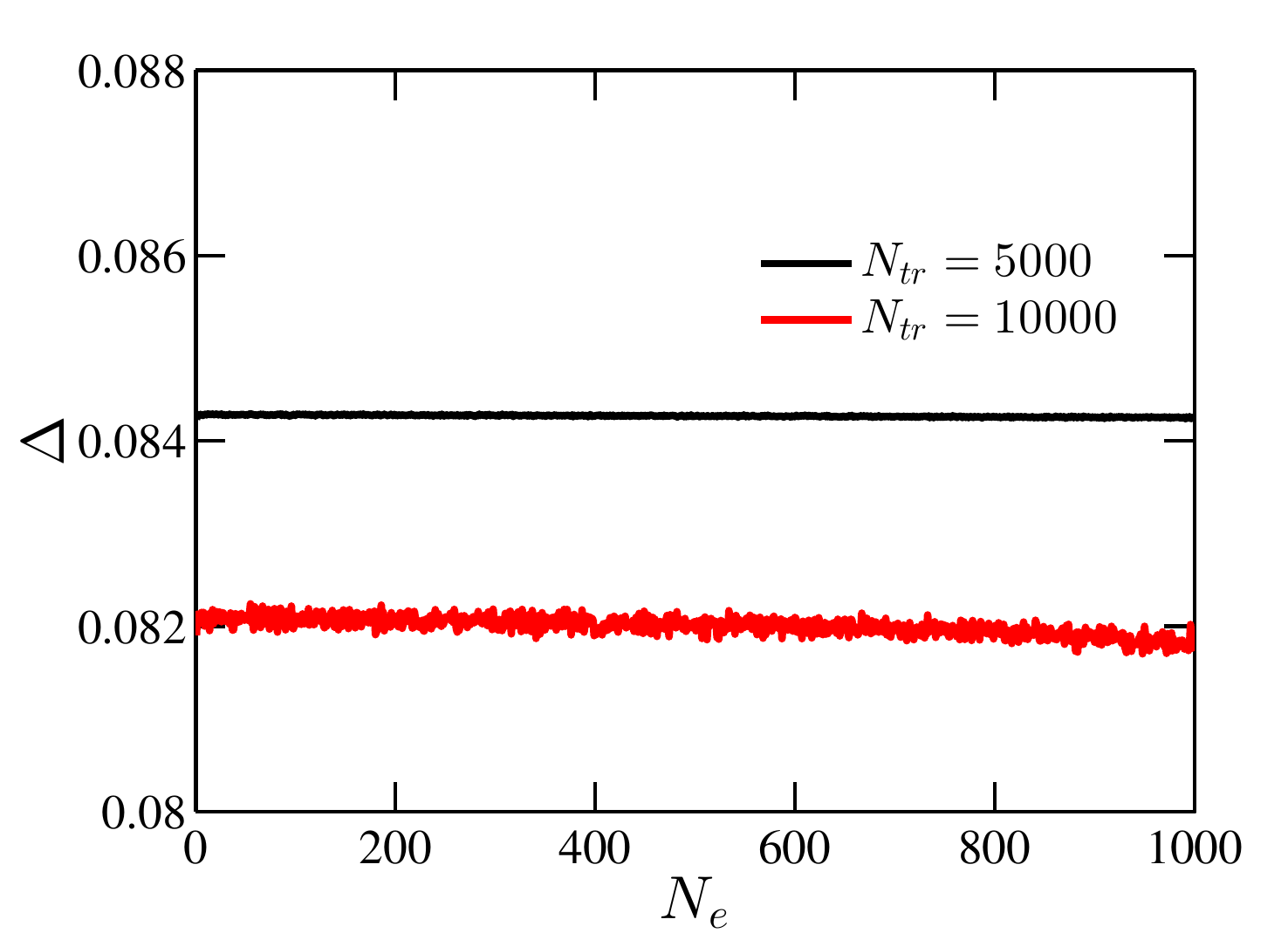}
	\caption{The averaged absolut error $\Delta$ for the exponent $\alpha$ as functions of the number of epochs. There are four hidden layers with each layer 25 neurons, namely, $N_{n}=25$. The black line corresponds to a training set with 5000 entries while the red/gray one with 10000 entries.}
	\label{fig:appen_alpha}
\end{figure}

\begin{figure}
	\includegraphics[width=0.85\columnwidth]{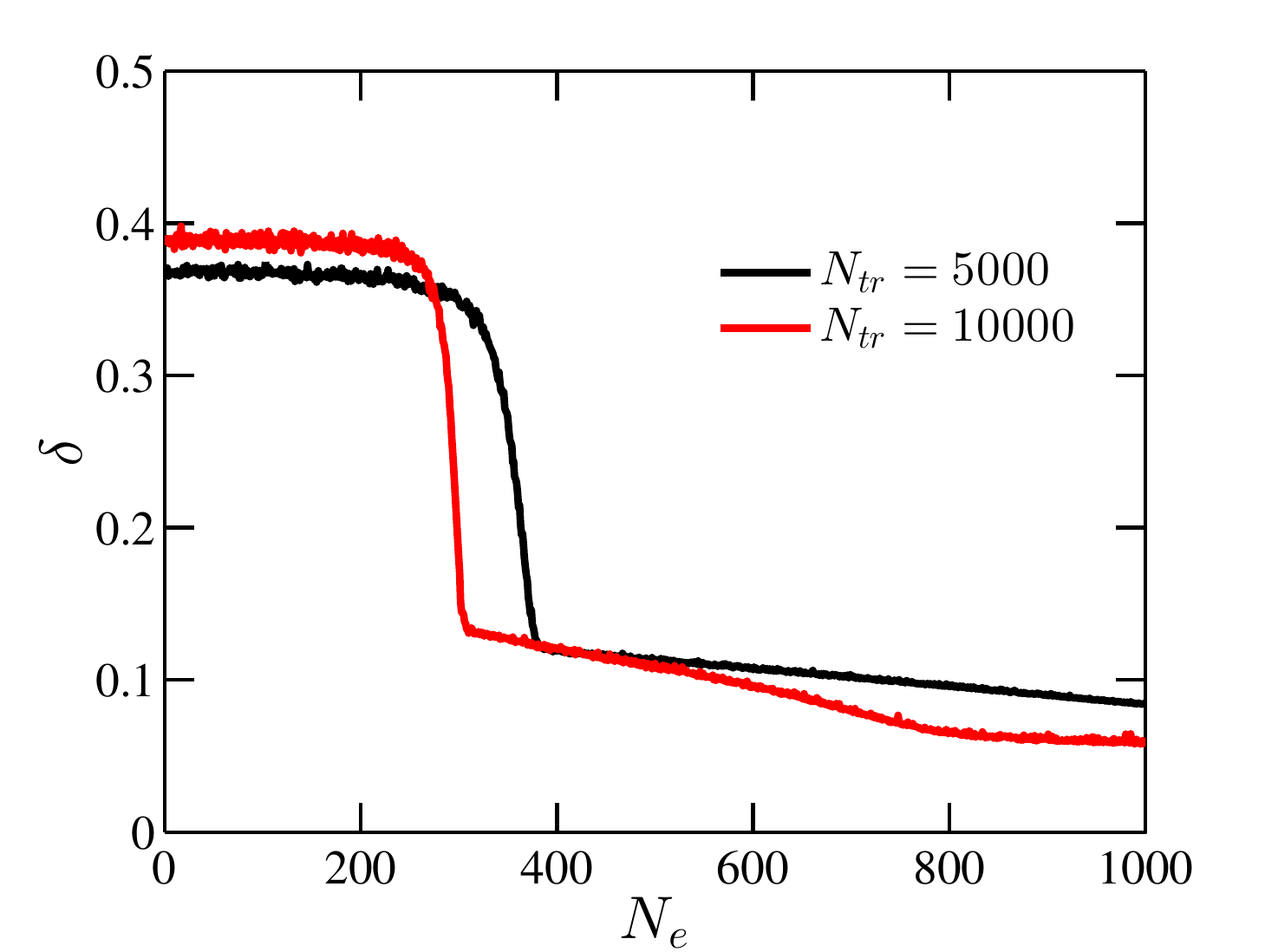}
	\caption{The averaged relative error $\delta$ for the noise amplitude $At_{0}$ as function of the number of epochs. There are four hidden layers with each layer 25 neurons, namely, $N_{n}=25$. The black line corresponds to a training set with 5000 entries while the red/gray one with 10000 entries.}
	\label{fig:appen_pd}
\end{figure} 

All results shown in the paper before this section are obtained by neural networks with two hidden layers (i.e. having a total of $L=4$ layers including the input and output ones). One may ask a question, would the results be better if we have more layers? In this section, we show some results obtained using networks with four hidden layers (i.e. $L=6$). We shall see that having more layers will not qualitatively improve the data. 

To ensure a fair comparison, we set the number of neurons in each layer $N_n=25$ for the network with four hidden layers, so that it has the same total number of neurons with those previously used ($N_n=50$ for two hidden layers). We have also checked results with more neurons, and no qualitative difference are found. We therefore focus on $N_n=25$ in this section. 

Fig.~\ref{fig:appen_alpha} shows the results for the averaged absolute error $\Delta$ of the exponent $\alpha$. The two lines are obtained from two training sets with different sizes. For $N_{\rm tr}=5000$ the absolute error converges slightly above 0.084 while for $N_{\rm tr}=10000$ the converged error is about 0.082. In the results shown in Fig.~\ref{fig:4} the averaged absolute error is about 0.05. Therefore having more hidden layers does not help improve the training result.

Fig.~\ref{fig:appen_pd} shows the relative error $\delta$ for the noise amplitude $At_{0}$. For $N_{\rm tr}=5000$ the error is reduced to approximately 0.09 after 1000 epochs of training, while with a larger training set ($N_{\rm tr}=10000$) the error is improved to slightly about 0.05. This result is again comparable to those shown in Fig.~\ref{fig:5}.

Overall, we have shown that having two more hidden layers would not offer additional improvement to the training outcome. While an exact formula determining the number of neurons and layers most suitable to a given problem is unclear, we speculate that this stems from the nature of the training process: there is always a competition of under-fitting and overfitting. If the complexity of the neural network (including number of neurons and layers) is lower than the complexity of the data/problem, the network tends to under-fit. Namely the network is less capable to produce the exact input-output correspondence given by the training set. On the other hand, if the complexity of the neural network is more than that of the data or problem, the network tends to over-fit, meaning that a lot of noisy details of the training data could appear in the training outcome. For many problems, two hidden layers are good enough.

\end{document}